\documentclass[conference]{IEEEtran}
\usepackage[T1]{fontenc}
\usepackage{cite}
\usepackage[utf8]{inputenc}
\usepackage{url}
\usepackage{amsmath,amssymb,amsfonts}
\usepackage{algorithmic}
\usepackage{graphicx}
\usepackage{textcomp}
\usepackage{xcolor}
\usepackage{setspace}
\usepackage{float}
\usepackage[bookmarks=false]{hyperref}
\usepackage{epstopdf}
\usepackage{grffile}
\usepackage{comment}

\def\BibTeX{{\rm B\kern-.05em{\sc i\kern-.025em b}\kern-.08em
    T\kern-.1667em\lower.7ex\hbox{E}\kern-.125emX}}
\begin{document}

\title{Spark-Based Port and Net Scan Detection}

\author{\IEEEauthorblockN{Antonia Affinito*, Alessio Botta*+, Luigi Gallo*, Mauro Garofalo*, Giorgio Ventre*+}
\IEEEauthorblockA{\textit{University of Napoli Federico II, NM2 SRL, Napoli, Italy} \\
\{antonia.affinito, alessio.botta, luigi.gallo3, mauro.garofalo, giorgio.ventre\}@unina.it
}}

\maketitle
\begin{abstract}
The two most spread network anomalies are port and net scan.
In this work, we present and analyze the results obtained by traditional approaches on Apache Spark. The use of Big Data technologies grants to significantly reduce the execution times of the algorithm, so to be used even in current high-speed networks.
The paper describes our approach and presents an experimental analysis in terms of detection performance and execution time. We use real traffic traces from MAWI archive and MAWILab anomaly detectors to compare with our results.
The analysis shows that i) our traditional threshold-based algorithm is already able to achieve detection performance higher than MAWILab (in 95\% of the considered cases with the best threshold value), currently considered the gold standard in the field; ii) the execution time is much shorter than the trace time, which makes it usable also in real time. Moreover, we bridge an important gap in literature providing the research community with a new labeled dataset, validated using MAWILab and extended with other anomalies not detected by it.
\end{abstract}


\section{Introduction and motivation}

The pervasive use of the Internet has led to a significant increase in the amount of traffic that crosses the network every day, so the amount of data that an anomaly detector has to analyze is higher and higher, especially in current high-speed networks.
Network traffic can be analyzed at several layers of the protocol stack: packet, flow, application, etc.
At flow-level, packets relating to the same TCP or UDP communication (e.g.~all packets related to an HTTP communication from and to a single host and a web server) are aggregated, and a summary of such group of packets is considered.
These summaries can now be provided directly by network devices such as switches and routers, and standard protocols have been defined for this aim (e.g.~Internet Protocol Flow Information Export or IPFIX~\cite{ipfix:online}).  Working at flow level seems the most promising approach for coping with the high-speed of current and future networks.
However, even at the flow-level, the detection of anomalies in high-speed networks require huge computational power or data reduction techniques as flow records represent a huge quantity of data.

In this paper, we analyze traffic traces from high-speed networks using a flow-level approach. The aim is to detect two of the most spread network anomalies i.e.~port scan and network scan (or simply net scan). In the former case, an attacker probes a host (the victim) on various TCP/UDP ports to find active and vulnerable services. In the latter case, the attacker scans a group of victim hosts on a single or a small number of ports. Such scanning activities are also associated with worms and botnets~\cite{zou2006}.

Port and net scans generate a real specific pattern in network traffic.
One of the most popular methods for detecting scanning activity is based on the fan-in fan-out proportion of the hosts, i.e.~ counting the number of incoming and outgoing flows, and comparing their ratio with a threshold~\cite{1705616, 1317747}.
With this approach, the performance in terms of detection capacity can be high, but the performance in terms of execution times can be very low ~\cite{5455789, 1705616, 1317747}. Sampling is typically applied to solve this problem, but this involves a significant loss of information. To overcome this problem, we have used Big Data analysis techniques to analyze a large amount of data with a threshold-based algorithm in the shortest possible time.
In particular, we use the Apache Spark framework (Spark in the following), which is comparable to Hadoop Map/Reduce but it provides faster results working entirely in the memory.
We implemented a simple threshold-based detection algorithm in Spark and tested it by using several real traces.
Our results in terms of execution time show that we achieve a processing rate up to 1/16 of the trace duration, which makes the approach able to run in real time.
Comparing our detection performance with MAWILab
~\footnote{Mawilab Documentation: \url{http://www.fukuda-lab.org/mawilab/documentation}}, we show that our approach achieves fewer false negatives than
MAWILab which is to say that we can uncover more anomalies than the gold standard.

Our contributions can be summarized as follows: i) we show that the threshold-based approach can achieve higher detection performance than the available gold standard based on much more complex and therefore less observable approaches; ii) we show that threshold-based approaches can be applied to current network traffic using Big Data technologies to achieve the required processing speed; iii) we provide an improved and constantly updated dataset which can be used by the research community for further studies on this important topic. The new dataset is automatically released on our SPADA project website \url{http://spada.comics.unina.it}.

We believe this represents an important step forward for the research community focused on port and net scans.

\subsection{Scanning Activities over Time}
\label{anomalies}
\begin{figure}[h]
\centering
\includegraphics[width=0.38\textwidth]{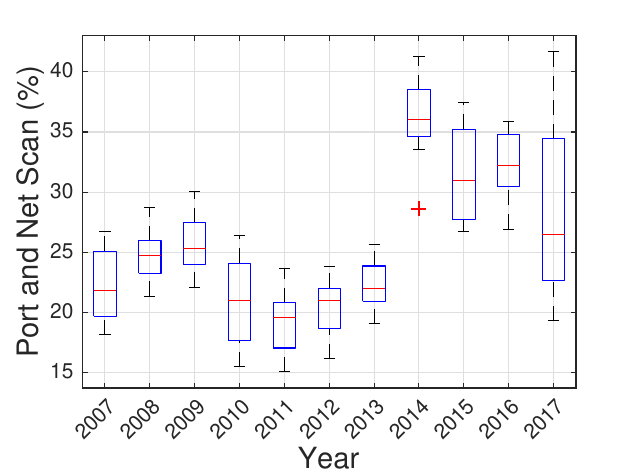}
\caption{Ratio of scanning over all anomalous activities over time during the last 11 years.}
\label{fig:anomaliesbox}
\end{figure}

Our first question in this work was if net and port scan anomalies are still actual today, so to justify further studies on this topic. To answer this question, we have conducted a longitudinal analysis of the number of anomalies detected by MAWILab over time.
Fig.~\ref{fig:anomaliesbox} shows the box plot of the scanning anomalies related to the years from 2007 to 2017
The image presents a growing trend, with a large increase in the ratio starting from 2014. This result indicates that the scanning anomalies are increasingly present in the traces: an increase that has been evident especially in recent years. This behavior motivates our choice to focus on these types of the anomaly and calls the research community for always updated data, techniques, and tools for port and net scan detection.

\section{State of the Art}
\label{sec:stateArt}
Scientific literature has been focusing on network and port scan for several years. Generally, these anomalies are examined considering that a source of scanning activity shows a very high number of outgoing connections~\cite{5455789}.
\textit{Zhao et al.}~\cite{1705616} also proposed an approach based on this consideration
To cope with high connection speeds (10-40 Gb/s), they proposed a standard hash-based flow sampling algorithm. 
\textit{Dainotti et al.}~\cite{dainotti2006} used wavelet to detect network anomalies and to precisely locate their position inside the traffic, while \textit{Balram et al.}~\cite{balram2008} proposed a technique based on packet count through neural networks.
\textit{Sridharan et al.}~\cite{sridharan2006} compared the performance of Snort and Bro on backbone traffic and propose a new approach based on sequential hypothesis testing that achieves higher performance than existing ones.
\textit{Kim et al.}~\cite{1317747} described a scanning activity in terms of traffic models. Traffic is analyzed at flow-level, but the scanning anomalies are detected by analysis of variations in network traffic models.
\textit{Chan et al.}~\cite{adfinald26:online} proposed two machine learning methods, useful for the constructing of models detecting network anomalies starting from past behavior.
The approach described by \textit{Wagner et al.}~\cite{1566205} is based on probabilistic measurement of entropy, used to indicate regularity in traffic of network flows.
Traffic models used in the three last works can be sensitive to changes in the type of traffic and network.
Threshold-based approaches have been widely and successfully used in the literature~\cite{1317747w}. In this work, we want to update these approaches to the current transmission rates and network technologies, and without linking the analysis to a specific point of the network.

MAWILab team proposed a system to detect attacks or anomalous events, applying a combination of four detectors with different theoretical backgrounds~\ref{sec:data}.
This system is currently used as a gold standard in literature~\cite{8071525}:
\textit{Casas et al.} proposed the combined use of a Big Data framework and machine learning algorithms to achieve high performance in terms of speed of execution and detection performance. They analyzed five types of anomalies.
We focus on the entire class of port and net scan and use a much simpler detection algorithm. Moreover, we uncover that MAWILab - the ground truth they, as many other works in literature, use - is incomplete. This clearly jeopardizes the results obtained. Therefore, we also propose an improved dataset, obtained through a combination of MAWILab and our algorithm.

\section{Our Approach}

\subsection{A simple detection algorithm}
\label{alg}

The algorithm used in this paper for port and net scan detection is described below.
Input data are flow-level information, where, in particular, we mainly concentrate on the timestamp, the IP addresses and the transport-layer ports of each flow.
For our experiments, we obtained the flow-level information processing the packet traces from MAWI with a tool named TIE (Traffic Identification Engine) ~\footnote{Traffic Identification Engine \url{http://tie.comics.unina.it}}. This tool combines the packets in flows using five fields: {Source IP Address, Destination IP Address, Source Port, Destination Port, Protocol}.
Then, our algorithm divides the flow-level trace into time intervals, denominated \textbf{slices}, of custom duration (e.g.~30 seconds).
Afterward, we use Spark SQL to calculate the proportion between generated and received flows in each slice and from each IP address. Then, this proportion is compared with a \textbf{threshold value}. The IPs whose proportion is larger than the threshold are marked as anomalous. 

It is worth noting that, even if it is very simple, this algorithm is still quite robust. For instance, it does not mark as anomalous hosts those generate a large amount of, even unbalanced, flows (e.g.~servers serving popular applications) because a few responses from the other hosts (e.g.~the clients) is sufficient to re-balance the equation.
As we will see in Sec.~\ref{sec:valid}, this simple algorithm is able to detect port and net scan anomalies with high precision and recall and in very short execution time if run on Apache Spark.
It is also worth specifying that this algorithm cannot detect other types of anomalies or more sophisticated attacks by design, but it does not require traffic sampling or modeling.
Moreover, it is able to detect all types of port and net scans unlike other works that focus only on specific types such as ~\cite{8071525} that analyzes only UDP and TCP-ACK net scan.
This algorithm is useful to show how the Big Data Analytics framework can solve long-unresolved issues in this important field of research. In fact, this algorithm, implemented on Apache Spark, achieves the high performance needed for detecting anomalies in large volumes of traffic, not recurring to a complex technique.


\subsection{Using Apache Spark}
\label{sec:spark}
Apache Spark is a platform for fast and efficient distributed processing of Big Data which has almost substituted Hadoop \footnote{Welcome to Apache Hadoop! \url{https://hadoop.apache.org}}. It is very fast both in storage and in data processing because it supports \textit{in-memory} processing, which allows analyzing data directly in main memory ~\cite{ViewofBi3:online}. 

Apache Spark can work in two different ways: \textbf{Batch} and \textbf{Streaming}.
Both modes~\footnote{Spark Streaming - Spark 2.3.0 Documentation. \url{https://spark.apache.org/docs/latest/streaming-programming-guide.html}} have been used in this work.
Apache Spark allows data storage in three different types of structure: Resilient Distributed Dataset (RDD), Dataframe, and Dataset. In this work, the DataFrame structure is used: it is conceptually equivalent to a table in a relational database
On these types of tables, it is possible to execute SQL queries using SQL commands, whose results are still a DataFrame.
Apache Spark supports different programming languages, e.g.~Java, Python, and Scala. We used Scala for two main reasons: i) Apache Spark is built on Scala and so if there are errors in the code or the source code does not have the expected result it is easier to debug; ii) Scala is about 10 times faster than others (Python) to analyze and to process data by the presence of the Java Virtual Machine.

\section{Data and methodology}
\label{sec:exp}

\subsection{Data used}
\label{sec:data}

We used several real traffic traces from the MAWI (Measurement and Analysis of the Wide Internet) dataset: an archive of real traffic traces provided by the MAWI Working Group~\footnote{Mawi working group traffic archive. \url{http://mawi.nezu.wide.ad.jp/mawi/}}. Traces are captured since 2007, and they constitute a very rich dataset that includes different applications and network conditions, including the presence of various known anomalies with global or local impact, periods of congestion, and network reconfiguration.
Traffic traces considered are captured on a transoceanic link that connects Japan and the United States of America. Each trace in the archive consists of traffic captured every day from 14:00 to 14:15 in different locations inside the WIDE network. Traces of 24 and 48 hours are also occasionally collected. A typical 15-minute trace is characterized by anonymized IP addresses, without payload, and contains 300k-500k unique IP addresses~\cite{7878657}. In this work, we use the traces captured at Samplepoint-F, a link working at 1 Gbps with a current average load of 650 Mbps that has largely increased in recent years~\cite{5061979}.

The MAWI group also created the MAWILab project, including a novel approach for the network anomaly detection, applied automatically every day on traffic from a specific MAWI repository. 
MAWILab defines a distance to normal traffic to recognize anomalies in MAWI traffic traces. This distance depends on the combination of four anomaly detectors based on different theoretical backgrounds: Principal Component Analysis (PCA), Gamma distribution, Kullback Leibler (KL) divergence, and Hough transformation. These detectors only work on the IP header~\cite{6906328, Fontugne:2010:MCD:1921168.1921179}.
The results of these detectors are combined to classify the anomalies in four types: \textbf{Anomalous} - assigned to all abnormal traffic and should be identified by any efficient anomaly detector, according to the authors; \textbf{Suspicious} -  assigned to all traffic that is probably anomalous but not clearly identified by their method; \textbf{Notice} -  assigned to all traffic that is not anomalous, but has been reported by at least one detector; \textbf{Benign} - all the rest of the traffic where no detector has labeled it as abnormal.

MAWILab provides the results of the analysis in two files, \textit{Anomalous} and \textit{Notice}, in which there are all the anomalies found. After detecting the anomalous behaviors, MAWILab applies a heuristic to assign a label related to the type of anomaly. Possible labels are represented in a tree-based taxonomy, where the root is a generic event and nodes contain an anomaly label.
In the first part of this work, we used MAWILab archive as a ground truth~\cite{Thegolds26:online} to validate our method (Sec.~\ref{sec:valid}). Successively, we verified that many anomalies were not detected by MAWILab, and built a new dataset on which we performed further analysis (Sec.~\ref{sec:valid3}).

It is worth noting that MAWILab database helped and still helps a lot of researchers to evaluate the performance of novel anomaly detectors. The availability of traffic traces is already scarce. Labeled traces, including an indication of anomalies inside them, are very very rare in our research community, and this is a great obstacle to further studies on this topic. For this reason, we decided to also evaluate MAWILab accuracy, and we finally managed to improve it. In particular, our dataset~\cite{spada} includes a larger set of port and net scans not detected by MAWILab, which we believe is an important contribution for the research community.

\subsection{Methodology}
\label{sec:valid}

The execution of the anomaly detection algorithm, described in Sec.~\ref{alg}, provides in output the IP addresses that are sources of port or net scans.
In the first part of our experimentation, we have used MAWILab as ground truth, i.e.~we compared the addresses, detected by our algorithm, with the ones in the \textit{anomalous} and \textit{notice} files provided by MAWILab. The IP addresses detected by our algorithm are also in one of these two files are considered true positives.
The results of this analysis are reported in Sec.~\ref{sec:valid2}.
We considered as false positives the ones detected by us and not by MAWILab and false negatives the ones detected by MAWILab and not by us.

For true and false positives we manually verified that all the IP addresses reported are actually anomalous.
Such analysis evidenced the limitations of MAWILab: several anomalies we detected were not present in both MAWILab files (i.e.~were even not considered suspicious by MAWILab). As explained in details in Sec.~\ref{sec:valid3}, we confirmed this result with several manual inspections of the pcap files. Starting from this important result, we constructed a new dataset extending MAWILab with other anomalous flows and used such dataset as a ground truth for another experimental analysis, reported in Sec.~\ref{sec:valid3}.
A \textbf{confusion matrix} is generated comparing the results of the threshold-based algorithm and the ones from MAWILab.
The detection performance of our anomaly detector is evaluated using two main metrics: \textbf{Recall}, also called True Positive Rate, and \textbf{Precision}. 

Several traffic traces have been analyzed in our experimentation. Results reported in the following refer to $60$ traces, characterized by a duration of 15 minutes and collected from December 2017 to September 2018. Similar results have been obtained on the other traces.
We have carried out multiple tests on different values of the threshold ranging from $20$ to $200$. A sensitivity analysis for this important parameter is reported in Sec.~\ref{sec:roc}. Most of the following results are then presented for the three threshold values that are more interesting: 50, 100, 200.\\

\section{Results of the experiments}

\subsection{Using MAWILab as a ground truth}
\label{sec:valid2}

In this section, we analyze the results we obtained using MAWILab as a ground truth. In particular, we used the Anomalous and Notice files from MAWILab and considered only scanning anomalies, which are the ones our detector has been designed for. All anomalies that are not part of scanning activities have been removed from MAWILab results (e.g.~normal events, Denial of Services, Distributed Denial of Services). Since the aggregation of packets into flows does not work well for ICMP, due to unbalanced reduction compared to TCP/UDP, ICMP anomalies are not considered.
\begin{figure}[t]
\centering
\includegraphics[width=0.34\textwidth]{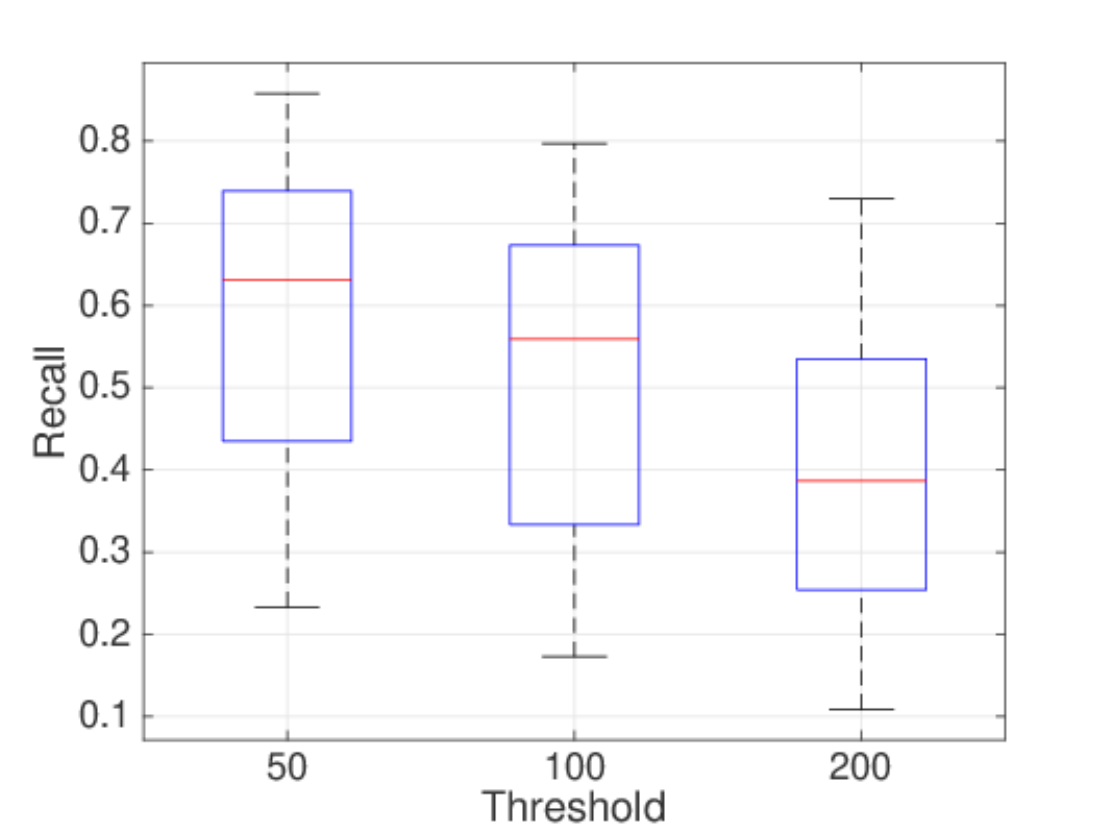}
\includegraphics[width=0.34\textwidth]{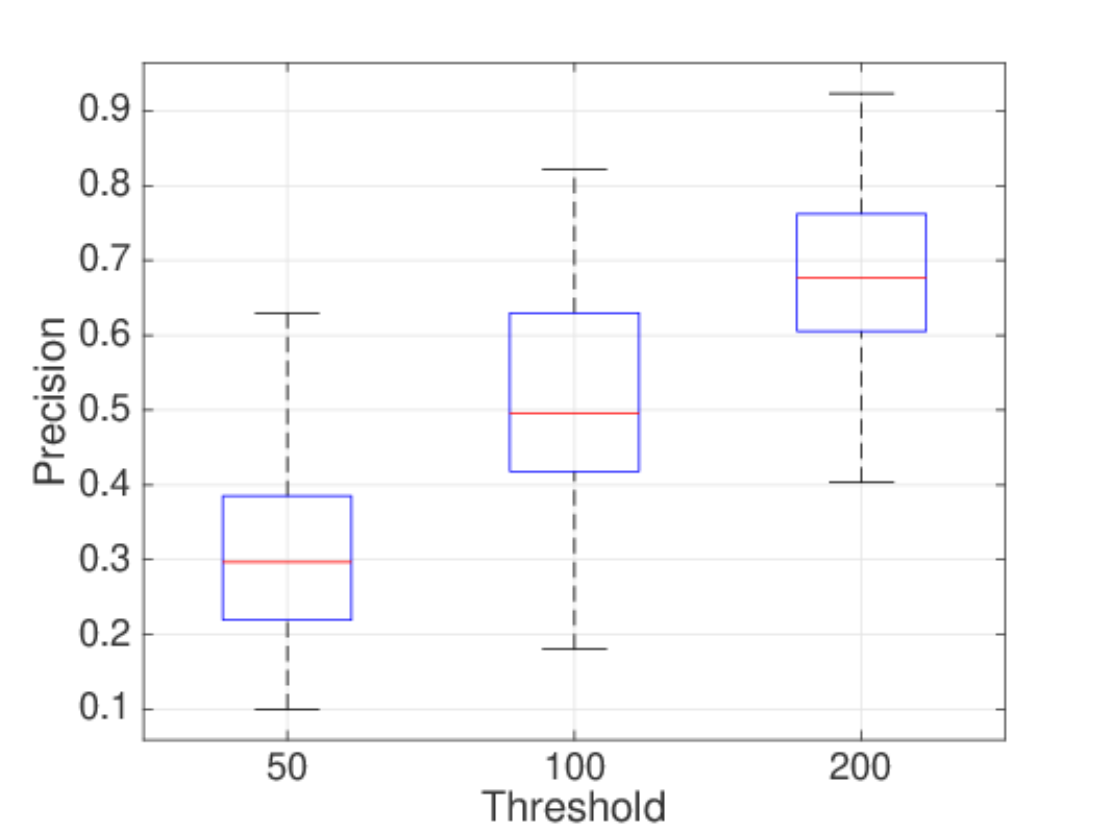}
\caption{Recall (left) and Precision (right) of the threshold-based approach using MAWILab as ground truth.} \label{fig:filtered_comparison}
\end{figure}
Fig.~\ref{fig:filtered_comparison} shows the values of the Recall and Precision obtained. In particular, we report the box plot of the precision and recall obtained for all the traces considered. Such figures illustrate the median value of the recall ranges from about $0.65$ to about $0.4$ increasing the threshold value. The median value of the precision ranges from $0.3$ to $0.65$ increasing the threshold value. Using a threshold value of $100$ we can obtain about $0.55$ for the recall and about $0.5$ for the precision. These results may seem to indicate that a threshold-based approach does not allow to obtain satisfactory results and cannot be used to detect such anomalies. We then analyzed the false positives in more details to validate this hypothesis.

We performed manual inspection of the pcap trace starting from the false positives. 
Such inspection revealed that the most part of the false positives was actually a source of scanning activity and MAWILab was unable to detect them.
For example, we noticed that several IP addresses generate flows with one or two packets, mostly with the TCP SYN and ACK flags set, and they receive zero or a very small number of answers. In addition, the number of useful bytes, i.e. bytes of the TCP payload, is usually zero. These considerations have led to consider them as IP addresses generating port and net scan.
We confirmed this result using several traces.
An important implication of this result is that we can not consider MAWILab as a ground truth as done up to now in different works in literature.

\subsection{Constructing and using the new ground truth}
\label{sec:valid3}

We built a new dataset expanding MAWILab with the IP addresses that have been found abnormal. In particular, based on the results of the previous analysis, we implemented two heuristic rules to complement MAWILab results for all the cases in which a source of the scanning activity was not detected by MAWILab. IP addresses that are false positives and trigger such rules are reintegrated into the true positives. The rules are the following: i) An IP address is a generator of \textbf{Net Scan} if it generates at least 20 flows towards different IPs of the same subnet; ii) An IP address is a generator of \textbf{Port Scan} if it contacts the same destination IP address on more than 10 ports.
Using these rules on several traces, we have built a new dataset to evaluate the performance of the threshold-based algorithm detector. This new dataset is obtained by the union of MAWILab results and the list of IP addresses that are considered a source of scanning activities after the application of the rules implemented.

\begin{figure}[h]
\centering
\includegraphics[width=0.24\textwidth]{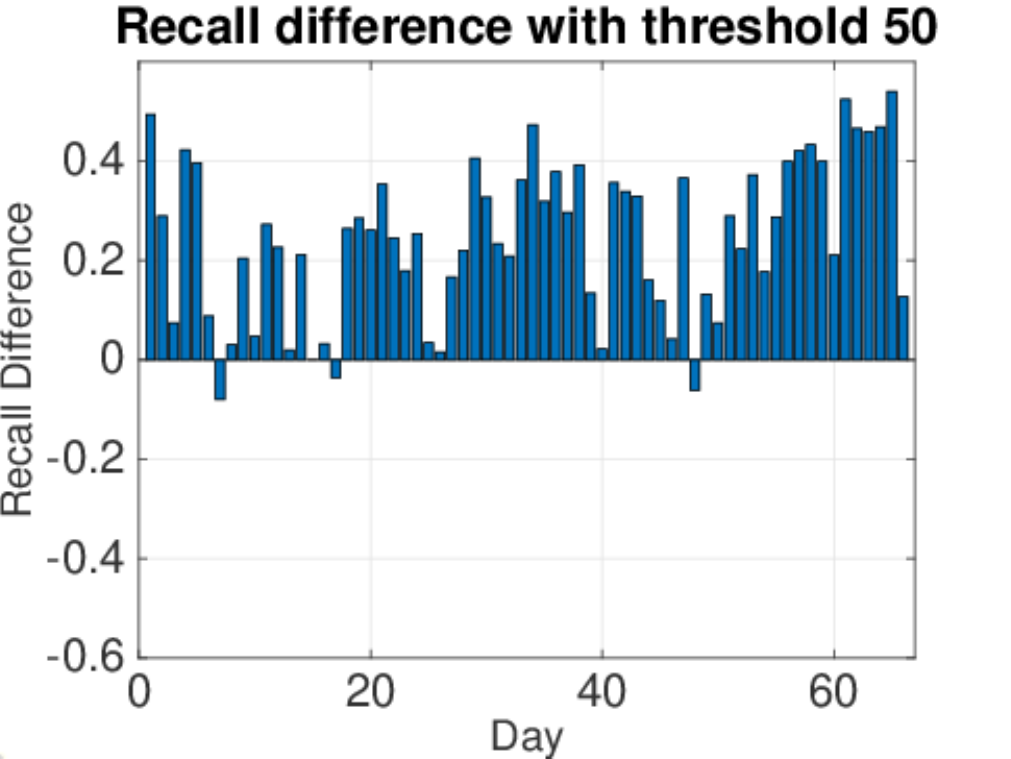}
\includegraphics[width=0.24\textwidth]{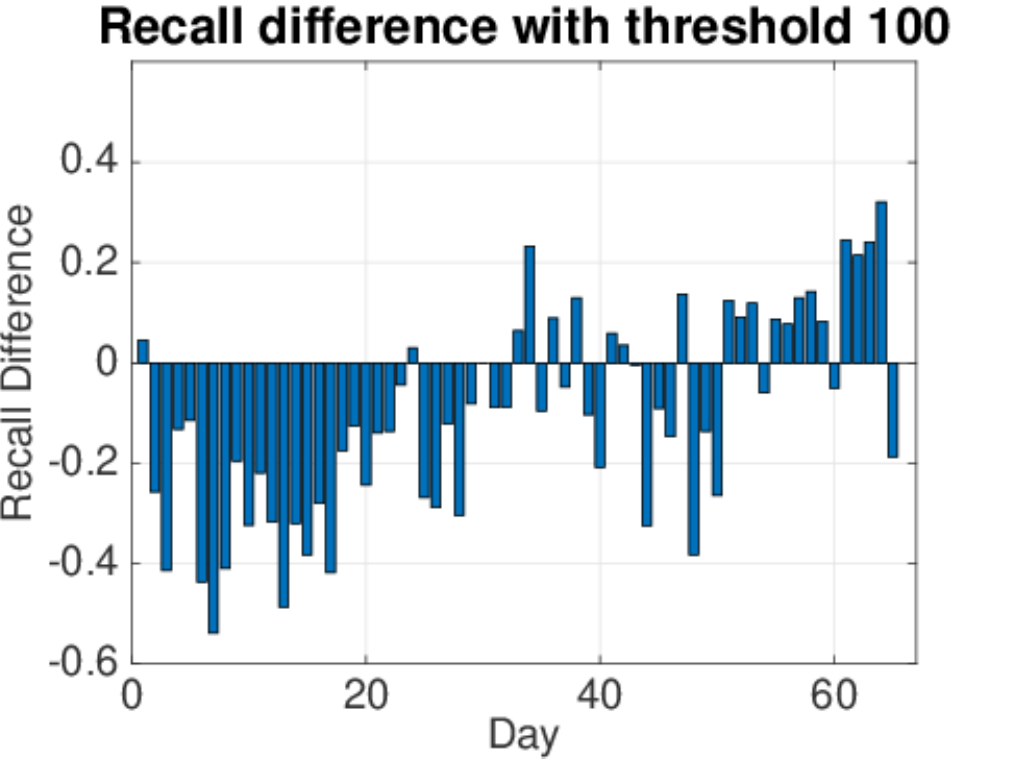}
\caption{Difference between the Recall of the threshold-based approach and the one of MAWILab using the new dataset as ground truth.}
\label{fig:realnnrecall}
\end{figure}
In the experiments described in the following, we compared MAWILab with the threshold-based approach and used the new dataset as a ground truth.
Fig.~\ref{fig:realnnrecall} shows the difference between the recall of the threshold-based algorithm and the one of MAWILab as a function of the different traces analyzed. The figure shows that the recall of the former algorithm is larger than the one of MAWILab in $95\%$ of the cases when the threshold value is $50$ and in $33\%$ of the case with a threshold value of $100$. When we increase the threshold value, MAWILab starts to obtain better performance in terms of Recall.

\begin{figure}[h]
\centering
\includegraphics[width=0.34\textwidth]{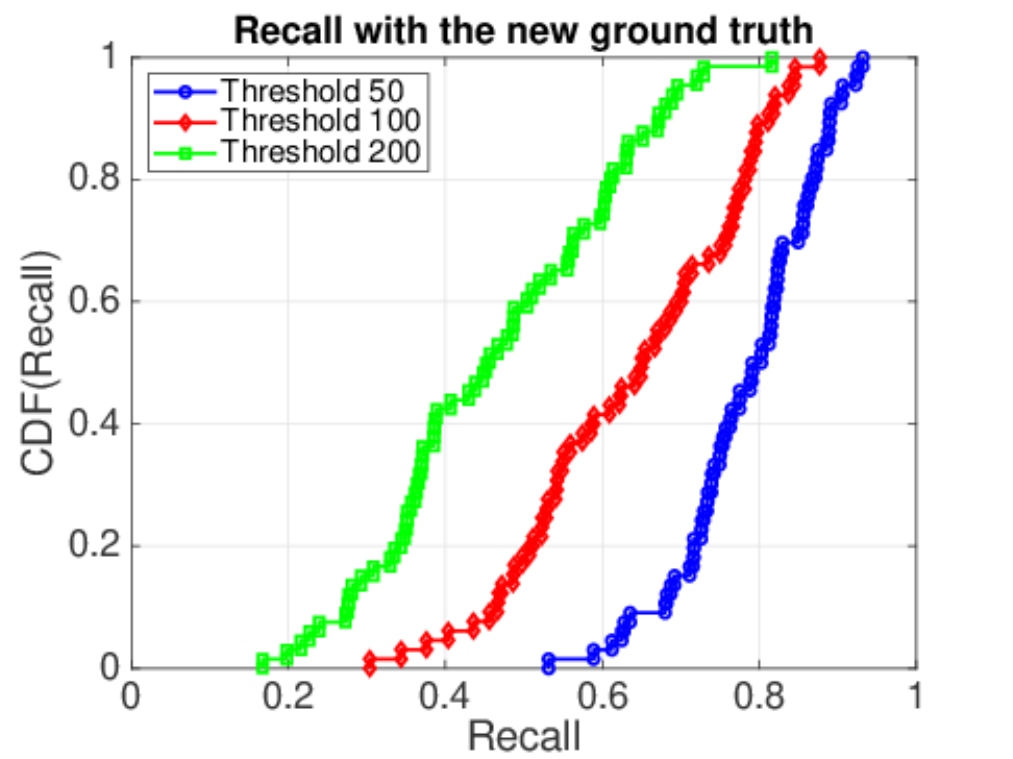}
\includegraphics[width=0.34\textwidth]{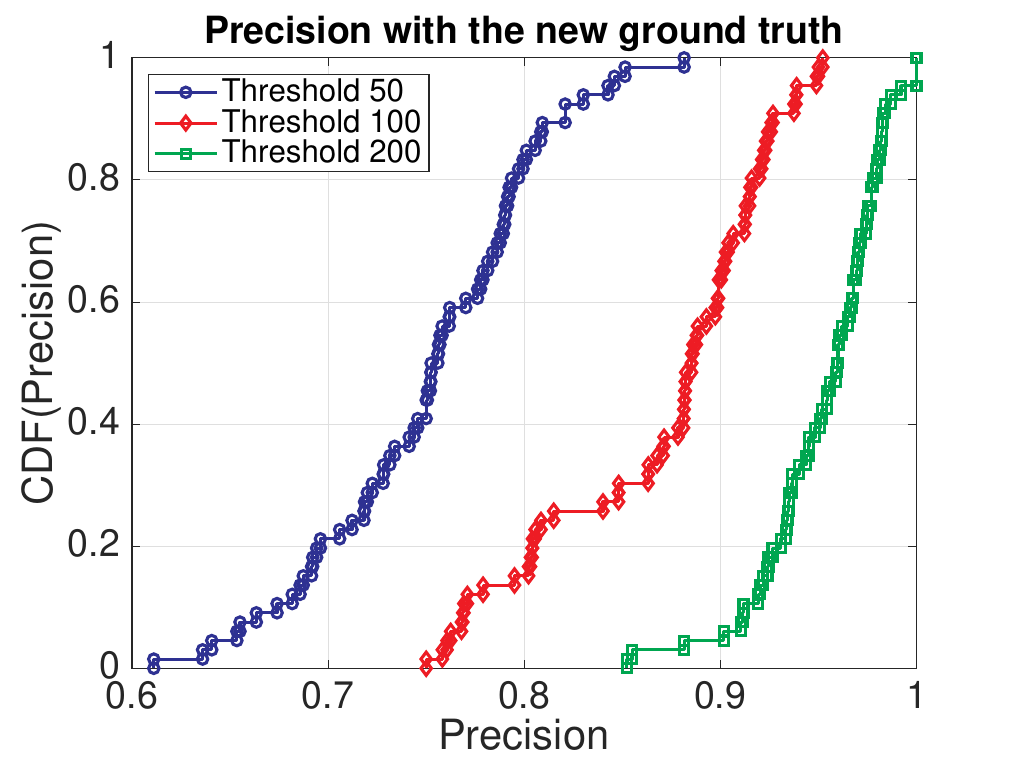}
\caption{Recall (a) and Precision (b) of the threshold-based approach using the new dataset as ground truth.}
\label{fig:realnnprecisiontot}
\end{figure}
The values of recall for the threshold-based algorithm are reported in Fig.~\ref{fig:realnnprecisiontot} (a).
The precision of MAWILab is clearly equal to 1 because there are not false positives.
Fig.~\ref{fig:realnnprecisiontot} (b) shows the precision of the threshold-based algorithm. We can see that for the intermediate threshold value (i.e.~$100$) the precision is larger than $0.85$ in about $70\%$ of the cases, and the median value is about $0.88$. Higher precision values can be obtained with higher threshold values.
Comparing the results in Fig.~\ref{fig:filtered_comparison} and  Fig.~\ref{fig:realnnprecisiontot} we can see the improvement of the performance of the threshold-based approach with the new dataset. Moreover, we can say that such very simple approach allows obtaining very high performance, higher than MAWILab which uses a much more complicated and therefore less observable approach.
Summarizing, in this section we have shown that very simple detection algorithm can obtain an even better Recall than MAWILab (in 95\% of the cases with a threshold 50 and in 33\% of the cases with a threshold 100), and this is because MAWILAb is not able to detect all scanning activities in MAWI traffic traces.

\subsection{Analysis of sensitivity to the threshold}
\label{sec:roc}

\begin{figure}[ht]
\centering
\includegraphics[width=0.35\textwidth]{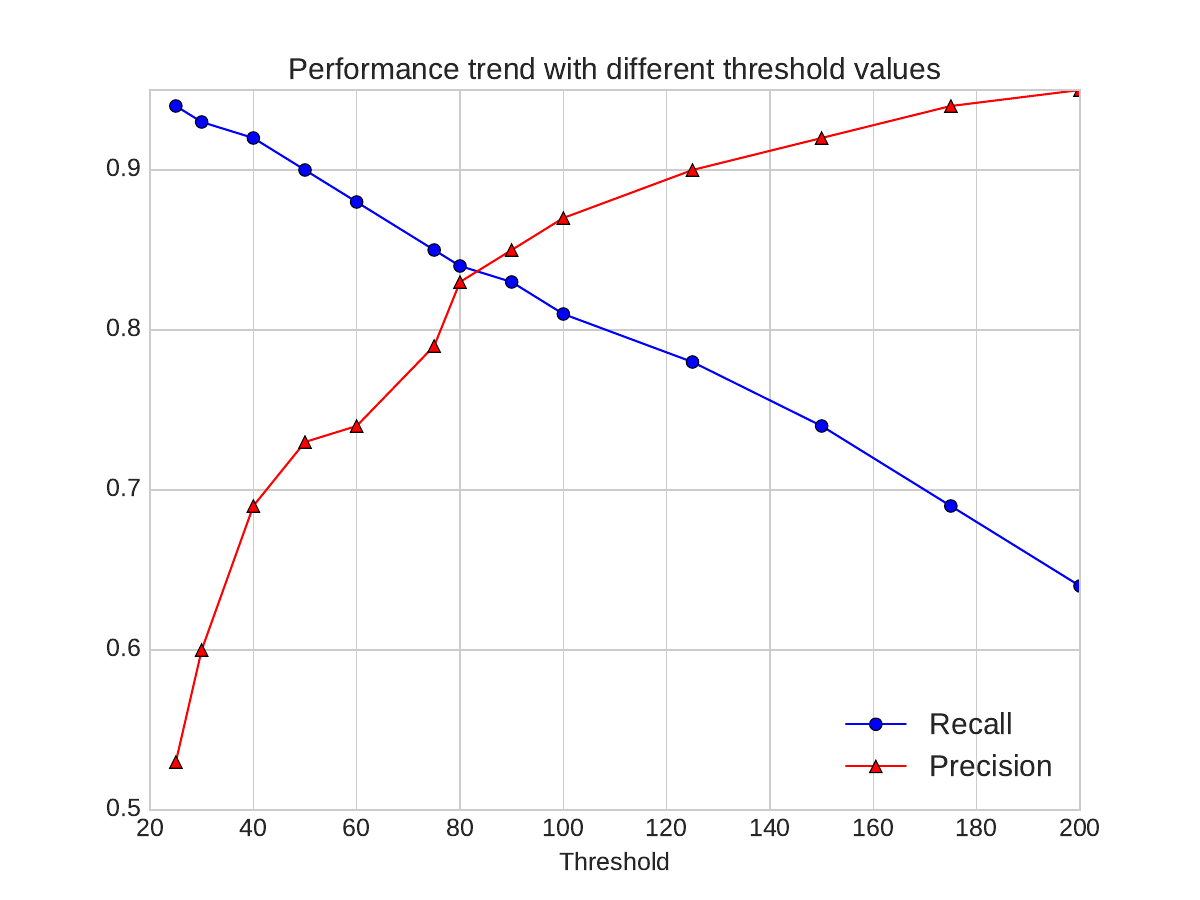}
\caption{Average Recall (blue) and Precision (red) using the new dataset as ground truth for several threshold values.} \label{fig:roc}
\end{figure}

Fig.~\ref{fig:roc} shows the average values of recall and precision obtained for several threshold values ranging from $20$ to $200$ using the new dataset as a ground truth. This figure shows that the best threshold value depends on which metric you want to maximize. For example, if false positives are annoying for a human intervention in the security pipeline, a good threshold value is about $125$. For such value, an average precision of about $0.9$ can be obtained. On the other side, if false negatives are more a problem, $50$ is the best threshold value to have a high recall without losing too much precision. An optimal value for both metrics is $80$, which allows obtaining an average recall and precision of about $0.85$.

\subsection{Speed analysis}
\label{sec:valid4}
\begin{figure}[h]
\centering
\includegraphics[width=0.38\textwidth]{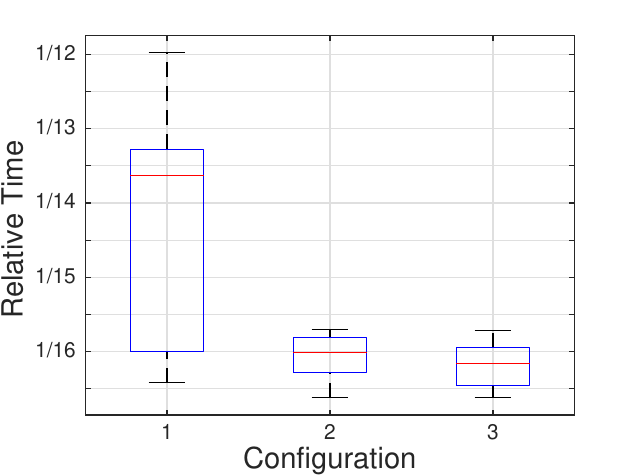}
\caption{Execution times of the threshold-based approach on Apache Spark in three different configurations.} \label{fig:boxplot}
\vspace{-10pt}
\end{figure}

In this section, we report the results of the analysis of execution times of the threshold-based algorithm.
Longer traffic traces, with a duration of 24 hours, have been taken from MAWI Dataset and used for this analysis.
Spark was used on virtual machines with different characteristics, all running on a private cloud.  In particular, we have instantiated a master machine with 8 GB of RAM and 4 cores and different worker instances:
\begin{itemize}
\item \textbf{First Configuration}: A machine with 32Gb of RAM and 4 cores.
\item \textbf{Second Configuration}: Two machines with 32Gb of RAM and 4 cores.
\item \textbf{Third Configuration}: A machine with 64Gb of RAM and 8 cores.
\end{itemize}
Fig.~\ref{fig:boxplot} shows the box plot of the execution times related to the three configurations. The x-axis represents the three configurations, and the y-axis represents the ratio between the execution time and the duration of the trace.
In the first configuration, the average execution time is about 1/14 of trace duration.
In the second and third configurations, i.r.~ more machines and more resources on the single machine respectively, the execution time is shorter than in the first one. In addition, there is no particular improvement between the second and the third configuration.


\section{Conclusion} \label{sec:conclusion}
Anomaly detection in network traffic is a hot topic in scientific literature, and it is in continuous evolution. Many proposals have been presented for this important research problem, but still today these kinds of anomalies are difficult to be detected efficiently and with high precision.
In this research area, two well known malicious activities are port and net scans.
In this paper we have firstly shown that these anomalies are still ingreasigly spread in recent years.
Several research efforts are currently put on complex techniques for anomaly detection, such as deep learning. These techniques can provide good detection performance~\cite{8071525}, but their observability is limited.
Our idea, instead, was to recover traditional detection approaches and resort to novel computing frameworks for obtaining the performance required by current network traffic.

In particular, we used a threshold-based algorithm, working at flow-level. The basic idea of this algorithm is to recognize malicious hosts looking at the ratio of their fan-in and fan-out (i.e.~the number of outgoing and incoming flows). Though very simple, this approach can obtain good detection performance, but it has scarce performance in terms of processing time. To cope with this problem, exacerbated in high-speed networks, we used Big Data Analytics and Apache Spark in particular.

We conducted an experimental analysis with several real traffic traces from the MAWI archive. We also used MAWILab anomaly detection results as a ground truth in the first part, and a comparison in the second one, after recognizing that MAWILab fails to detect several scans.
We firstly evaluated the precision and recall of the threshold-based algorithm using MAWILab as a ground truth.
Results were not satisfactory, in particular in terms of false positives, and pushed us to investigate more in deep.
Through manual inspections of several traffic traces, we verified that such positives were actually true rather than false. 
This drove us to create a new dataset starting from MAWILab and complementing it with several other anomalies identified through heuristic rules devised thanks to the analyses described above.
We then evaluated the performance of the threshold-based algorithm using the new dataset as ground truth and compared obtained results with the ones from MAWILab. This analysis shows that the simpler algorithm can easily achieve higher recall than MAWILab, which is already based on much more complex algorithms.

Moreover, we also have shown that the threshold-based algorithm presents short execution times, varying between 1/12 and 1/16 of the duration of the trace, thanks to the adoption of Apache Spark.
Finally, we also set up a system that processes the traces available from MAWI every day and publishes a new dataset, including a comparison with MAWILab. We believe this was an important yet missing contribution for further studies on this research topic.

\bibliography{main}{}
\bibliographystyle{IEEEtran}

\end{document}